\documentclass[showpacs,twocolumn,amssymb,prb,aps,psf]{revtex4}%Physical Review B
\usepackage{longtable,graphicx,epsfig}

\begin{document}
\title
{X-ray study of the liquid potassium surface: structure and
capillary wave excitations}

\author{Oleg~Shpyrko}
\affiliation{Department of Physics, Harvard University, Cambridge
MA 02138 (U. S. A.)}

\author{Patrick~Huber}
\affiliation{Department of Physics, Harvard University, Cambridge
MA 02138 (U. S. A.)}

\author{Peter~Pershan}
\affiliation{Department of Physics, Harvard University, Cambridge
MA 02138 (U. S. A.)}

\author{Ben~Ocko}
\affiliation{Department of Physics, Brookhaven National Lab, Upton
NY 11973 (U. S. A.)}

\author{Holger~Tostmann}
\affiliation{Department of Materials Science and Engineering,
University of Florida, Gainesville, FL 32611}

\author{Alexei~Grigoriev}
\affiliation{Department of Physics, Harvard University, Cambridge
MA 02138 (U. S. A.)}

\author{Moshe~Deutsch}
\affiliation{Department of Physics, Bar-Ilan University, Ramat-Gan
52900 (Israel)}

\date{\today}

\begin{abstract}

\renewcommand{\baselinestretch}{1.3}
\noindent
We present x-ray reflectivity and diffuse
scattering measurements from the liquid surface of pure potassium. They
strongly suggest the existence of atomic layering at the free surface of a pure liquid metal
with low surface tension. Prior to this study,
layering was observed only for metals
like Ga, In and Hg, the surface tensions of which are 5-7 fold higher than
that of potassium, and hence closer to inducing an ideal "hard wall"
boundary condition.
The experimental
result requires quantitative analysis of the contribution to the
surface scattering from thermally excited capillary waves.
Our measurements confirm the predicted form for the differential
cross section for diffuse scattering,
$d\sigma /d\Omega  \sim 1/q_{xy}^{2-\eta}$ where
$\eta = k_BT q_z^2/2\pi \gamma $, over a range of $\eta$ and $q_{xy}$
that is larger than any previous measurement.
The partial measure of the surface structure
factor that we obtained agrees with computer simulations and
theoretical predictions.
\end{abstract}

% 61.25.Mv  liquid metals and liquid alloys
% 68.10.--m fluid surfaces and fluid--fluid interfaces
% 61.10.--i x-ray determination of structures

\maketitle

\subsection*{1. Introduction}

\renewcommand{\baselinestretch}{1.3}

In contrast to measurements of dielectric liquids,
which show a monotonically varying density profile at the
liquid-vapor interface, liquid metals exhibit a phenomenon
known as surface-induced layering \cite{Rice74, Evans81}. This had been thought to occur because
the Coulomb interactions between two constituents of the liquid metal,
the free electron Fermi gas \cite{March90}
and the classical gas of positively charged ions, suppress local surface
fluctuations that would otherwise conceal the layered atomic ordering that
occurs at hard walls \cite{Iwa92}.
On the other hand recent molecular simulations by Chacon et al. \cite{Velasco02}
argue that surface layering should occur in any liquid for
which the ratio of melting temperature $T_m$ and critical temperature $T_c$, $T_m/T_c$, is sufficiently small.
Surface-induced layering for liquid metals was first
 theoretically predicted by Rice et al. \cite{Rice74},
later confirmed by density functional calculations \cite{Evans81, Iwa92}
and computer simulations \cite{Harris87}, and more recently observed
experimentally in a number of pure liquid metallic systems such
as Hg \cite{Mag95}, Ga \cite{Regan95}, In \cite{Tost99} as well as in binary alloys \cite{Tost98, Dimasi00}.
In fact, for Ga, Hg and K   $T_m/T_c \approx 0.13-0.15$ and these
observations are consistent with Chacon et al. \cite{Velasco02}.
From another point of view, all but one of the metallic systems studied so far exhibit
relatively high values of surface tension, a few hundred mN/m,
leaving unanswered the fundamental
question of whether the surface-induced layering is caused by
the high surface tension, which suppresses the long wavelength capillary
waves, or whether it is caused by the intrinsic metallic properties of the studied
systems. In the later case surface-induced layering should also be found
in low-surface tension metals.
Monovalent alkali metals have a surface tension comparable to that of water (72 mN/m) and their study should, therefore, shed light on this issue.
Moreover, monovalent alkali metals are the best
available examples of an ideal nearly-free-electron metal \cite{Ash76},
therefore, they are particularly well
suited for studying surface induced
layering in an ideal, simple liquid
metal. The same properties which make alkali
metals ideal for studies of layering have
also been the motivation for the numerous
computer simulations of liquid Na, K and Cs \cite{Rice74, Evans81, Harris87, Ash76, Has82, Chac84, Gomez94, Good83}.

Until recently, experimental surface studies of the pure liquid alkalis
were thought to be impossible because of two fundamental experimental
problems. First, their high evaporation rates which lead to high values
of vapor pressure \cite{Iida93}, and high reactivity \cite{Borg87} to water
and oxygen implied that their surfaces could not be maintained atomically
clean for extended periods of time even under UHV conditions. Therefore, the
first study of alkali metal systems performed by our group \cite{Tost99} employed a
binary KNa alloy that preserves most of the characteristic features typical
of a nearly-free-electron metal, but for which the vapor pressure and
melting temperature were both lower than those of the pure metals. Contrary
to expectations, no measurable changes were detected in that study in the
x-ray reflectivity of the alloy's surface over periods of many hours or
after intentional exposure to large doses of oxygen. This lack of surface
contamination is believed to be due to the high solubility of the oxides of
these metals in the bulk liquid \cite{Borg87, Add84}, and to the low surface tension of
liquid alkali metals which does not promote segregation of the oxide at the
surface. The fact that the surface of the KNa alloy was found to remain
atomically clean for extended periods of time suggests that the same might
be true for the pure metals. The second experimental problem is that the
large capillary-wave-induced roughness due to the low surface tension
renders the x-ray reflectivity peak weak and inseparable from its'
diffuse wings. Thus, a clear layering peak in the measured reflectivity
curve, such as found in the high-surface-tension liquid metal studies \cite{Mag95, Regan95, Tost99, Tost98, Dimasi00},
can not be observed here. We demonstrate here that this problem can be
resolved by a subtle application of diffuse scatttering and reflectivity
measurements. A significant aspect of this paper
pertains, therefore, to the
methods of measurement and analysis that are necessary in order to separate
the effects due to the thermal capillary waves from those of the
surface-induced layering. Using these methods, the present study clearly
demonstrates the existence of surface induced layering in liquid K, with
properties similar to those of the high-surface-tension liquid metals
studied previously \cite{Mag95, Regan95, Tost99, Tost98, Dimasi00}.

\subsection*{2. Experimental Details}

The sample preparation procedure and the experimental setup were similar to our previous
experiment on the KNa alloy \cite{Tost99}.
The sample was prepared in an Argon-filled glove box ( $ < $ 2ppm $H_2$0, $ < $ 1
ppm $O_2$). The K with purity of 99.999 $\%$, which was contained
in a glass ampoule, was melted
and then transferred into a sealed stainless steel reservoir
inside the glove box using a glass syringe.
The reservoir was sealed with a Teflon o-ring and then mounted onto
a UHV chamber.

\begin{figure}[!]

\label{fig:1}
\begin{center}
\unitlength1cm
\begin{minipage}{8.5cm}
\epsfxsize=8.5cm \epsfysize=8cm \epsfbox{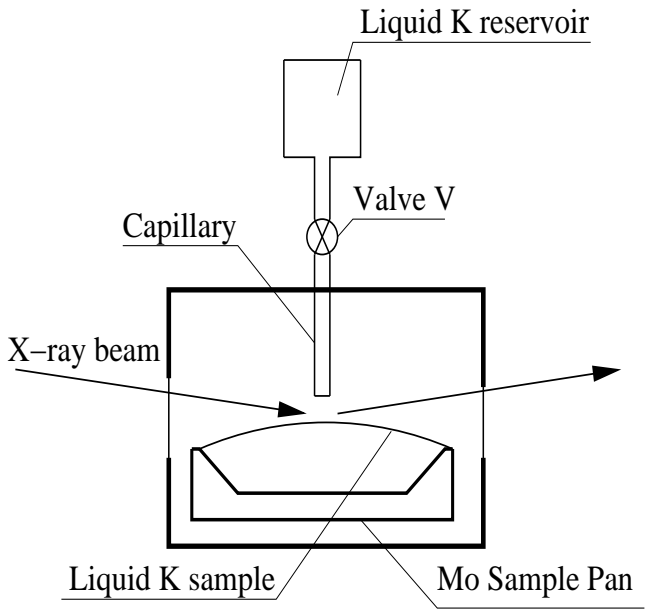}
\end{minipage}
\caption{ Schematic description of the experimental setup. }
\end{center}
\end{figure}

After a 24-hour bakeout of the entire chamber to 150
$^{\circ}$\,C, a base pressure of $10^{-9}$\,Torr has been
established inside the UHV chamber. The K in the reservoir was
then heated above the melting temperature $T_m$=63$^{\circ}$\,C
and dropped through a capillary into a Mo sample pan mounted
inside the UHV chamber by opening the valve V shown on Figure 1.
Previous to filling the pan with liquid K, the pan was cleaned by
sputtering with $Ar^{+}$ ions for several hours in order to remove
the native Mo oxide. This latter procedure significantly improves
the wetting of the sample pan by liquid K and therefore helps to
achieve a larger area of macroscopically flat liquid surface than
it has been possible for high surface tension metals
\cite{Regan95}. The sample pan was kept at a constant temperature
of 70 $^{\circ}$\,C. Pure K has a vapor pressure of about
$10^{-6}$\,Torr at this temperature, which renders the requested
UHV conditions almost impossible.  We circumvented this problem by
a differential pumping setup, where we used a permanently attached
ion pump. By constantly pumping at the sample chamber, we were
able to achieve a base pressure of about $10^{-9}$\,Torr in the
chamber, three orders of magnitude lower than vapor pressure of K.

\begin{figure}[!]
\label{fig:2}
\begin{center}
\unitlength1cm
\begin{minipage}{8.5cm}
\epsfxsize=8.5cm \epsfysize=8cm \epsfbox{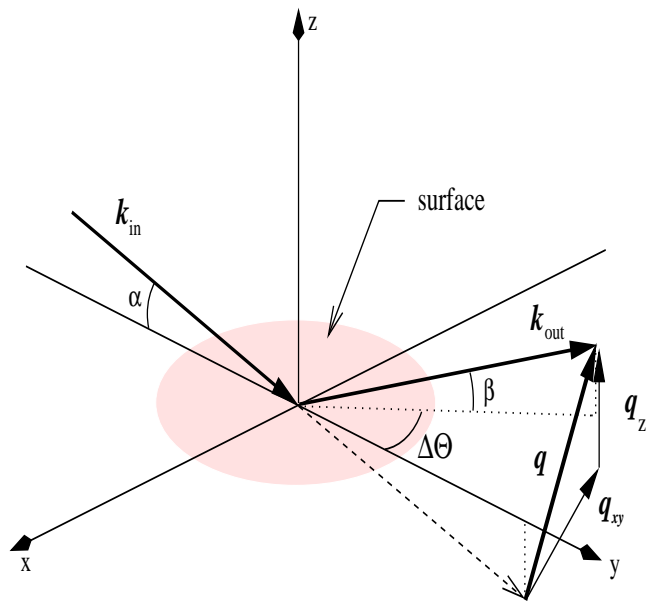}
\end{minipage}
\caption{ Kinematics of the x-ray measurement. }
\end{center}
\end{figure}

Measurements were carried out using the liquid surface
reflectometer at CMC-CAT beamline at the Advanced Photon Source,
Argonne National Laboratory, operating at an x-ray wavelength of
1.531 \,${\rm \AA}^{-1}$. The kinematics of the measurement is
illustrated in Figure 2. Two different surface-sensitive x-ray
techniques were applied - specular x-ray reflectivity and x-ray
diffuse scattering. In the specular x-ray reflectivity geometry,
the incident angle $\alpha$ and the reflection angle
$\beta$=$\alpha$ are varied simultaneously while keeping
$\Delta\Theta$ = 0 constant. The signal is measured as a function
of the surface-normal momentum transfer $q_z = 2 \pi / \lambda
\left( \sin \alpha + \sin \beta \right)$ with the in-plane
momentum transfer $q_{xy}= 0$.

In the x-ray diffuse scattering geometry, the detection angle $\beta$
is varied within the plane of incidence $\Delta\Theta$ = 0, while the incident
angle $\alpha$ stays fixed. The observed scattering therefore follows
a trajectory in the ($q_x, q_z$) plane with $dq_z / dq_x =\tan \beta$
and $q_y$=0. The signals measured in the two geometries are, however, related
since the signal measured at $\beta=\alpha$ in the diffuse measurement
is the specular signal.

The differential cross-section of the
the diffuse scattering signal is given by \cite{Pershan00}:

\begin{equation}
\frac{d\sigma }{d\Omega } = \frac{A_{0}}{\sin \alpha }
\left( \frac{q_c}{2} \right) ^{4}\frac{k_B T}{16\pi ^2 \gamma }
\mid \Phi (q_z) \mid ^2
\frac{1}{q_{xy}^{2-\eta}} \left( \frac{1}{q_{\mbox{\scriptsize max}}}
\right) ^\eta \:
\label{eq:master}
\end{equation}
where $q_c$ \cite{AlsNielsen01} is the critical angle for total
external reflection of x-rays, $\gamma$ is the surface tension,
 $q_{max} \approx \pi/\xi$
is the value of the upper cutoff for capillary wave contributions
\cite{Pershan00}. The value for $\xi$ is empirically taken to be
the nearest neighbor atomic distance in the bulk
\cite{AlsNielsen01}. The scattering lineshape exponent $\eta$ is
given by \cite{Pershan00}

\begin{equation}
\eta = \frac{k_BT}{2\pi \gamma} q_z^2
\label{eq:eta}
\end{equation}
The principal goal of the present surface x-ray scattering
measurements is to determine the surface structure
factor \cite{Regan95}

\begin{equation}
\Phi (q_z) = \frac{1}{\rho_{\infty}} \int dz\frac{\left< d\rho(z) \right> }{dz}
\exp(\imath q_z z)
\label{eq:structure}
\end{equation}
which is given by the Fourier transform of the ($x,y$)-averaged intrinsic density profile
$ \left< \rho(z) \right> $
along the surface normal $z$ in the absence of capillary waves.
In this equation, $\rho_{\infty}$ is the bulk electron density.
The fact that the peak of the differential cross section at the
specular condition is algebraic, rather than a $\delta$-function
complicates the determination of $\Phi(q_z)$ from the measured
intensity. This can be done, however, when both specular reflectivity
and diffuse scattering measurements are available.

\subsection*{Practical Considerations}
It follows from Eq. 1 that the nominal specular reflectivity signal,
observed at $\alpha=\beta$ and $\Delta\Theta=0$ corresponds to the integral of the
$\left( 1 / q_{xy} \right) ^{2-\eta}$ over the area
$\Delta q_{xy}^{res}\left( q_z \right) = \Delta q_{x}^{res} \times \Delta q_{y}^{res}$ which is
the projection of the instrumental resolution function on the horizontal plane of
the liquid surface.
The $q_z$-dependence of $\Delta q_{xy}^{res}\left( q_z \right)$ is due
to the increase of the projected width of the detector resolution
$\mid \Delta q_y^{res} \mid=(2 \pi / \lambda) \sin \beta \Delta \beta $
on the plane of the surface as $\beta$ increases. The integration
results in the reflectivity scaling as
$\left(\Delta q_{y}^{res} / q_{max} \right)^\eta / \eta $
and since $\eta$ scales as $q_z^2$, the net effect is a Debye-Waller-like
decrease in the reflectivity as $q_z$ increases.
For example, if one defines a mean square surface roughness by
taking an average over a length scale $ \approx (\Delta q_{y}^{res})^{-1}$,
this roughness can be written as \cite{Schwartz90}
\begin{equation}
\sigma^2 (q_z) = (k_b T / 2 \pi \gamma) \ln (q_{max}/\Delta q_y^{res})
\label{eq:sigma}
\end{equation}

From this one can show that the reflectivity is proportional
to a Debye-Waller factor $\exp [-q_z^2 \sigma^2 (q_z)]$.
The important point to note is that for liquid surfaces
the signal measured at the specular condition
$q_{xy}=0$ depends on the resolution of the reflectometer.

The validity of the simple capillary wave model in the high $q$-range
remains controversial and untested.  Although the model assumes
that the energy for surface excitations
can be described by the capillary form
$\approx \gamma /q_{xy}^2 $ with a constant surface tension, $\gamma$,
over the entire range of length scales from the macroscopic
to a microscopic length of the order of
$\xi = \pi/q_{max}$, where $\xi$ is normally taken to be of the order of
the interatomic/molecular spacing $a$,
this is certainly not strictly true for lengths that are shorter than
a length that we might designate as $\xi_{cutoff} = \pi/q_{cutoff} > a$.
Nevertheless, the basic physics is not changed if we replace $q_{max}$
by the smaller $q_{cutoff}$ since the differences engineered by this are
really just a matter of accounting, in the sense that it determines the
relative contributions between the capillary wave and structure factor terms given in Eq. 1.
Nevertheless this is not a serious objection to the present
treatment, since the capillary wave model is certainly accurate
for values of $q_{xy}$ covered by the present measurements.
Even if the model were to fail for values of $q_{xy} > q_{cutoff} $, the shorter wavelength fluctuations  could be incorporated
into  the definition of a $q_{cutoff}$-dependent structure factor
$\Phi(q_z)$.
On the other hand, the fact that the intensity of off-specular
diffuse scattering predicted by Eq. 1 has been shown to be in
satisfactory quantitative agreement with measurements  for
a number of liquids over a wide range of $q_y$ and $q_z$ implies
that $q_{cutoff}$ is reasonably close to $q_{max} = \pi / a$ and the extracted
structure factor $\Phi(q_z)$ is a reasonable measure of the local profile.

There are three practical considerations involved in these measurements.

The first is that it is necessary to insure that the measurement
corresponding to the integral
over the resolution does not include a diffuse scattering contribution
from sources other than the surface. For flat surfaces, for which
the specular reflectivity can be described by a $\delta$-function
at $q_{xy}$=0, this can be accomplished
by subtracting the scattering measured
at some finite value of $q_{xy}$ from the signal measured at $q_{xy}$=0.
For liquid surfaces this separation is complicated by the fact
that the signal measured at finite $q_{xy}$ includes scattering
from the $1/q_{xy}^{2-\eta}$ tails of the cross-section. We circumvent
this problem in the present study by comparing the difference between
measurements at $q_{xy}$ =0 and at some $q_{xy}\neq 0$ with the
same difference as predicted by the theoretical differential
cross section (Eq. 1). Thus, we present all data in
this study as differences between a signal measured
at $\lbrace \alpha, \beta, 0 \rbrace$ and at $\lbrace \alpha,\beta,
\Delta \Theta_{offset}$ = 0.2 degrees$\rbrace$.
For specular reflectivity $\alpha=\beta$; however, for off-specular diffuse measurements $\beta \not= \alpha$.
Diffuse scattering from all
sources other than the surface are essentially constant over this range of
$\Delta\Theta_{offset}$. This is because these other sources of scattering,
such as diffuse scattering from either the bulk liquid or from the vapor above the surface,
only depend on the absolute value of the total momentum transfer
$\mid q \mid$ and this is essentially unchanged for small offsets
in $\Delta\Theta_{offset}$.

The second practical consequence of the $1 / q_{xy}^{2-\eta}$
form is that as $\eta$ increases it becomes increasingly difficult
to distinguish the singular specular peak from the off-specular
power-law wings. At $\eta \geq 2$ this distinction
is no longer possible \cite{Pershan00}, even with infinitely sharp resolution.
Unfortunately, for liquid K, $\eta$ reaches the limit of 2
at $q_z$=1.74\,${\rm \AA}^{-1}$, which is comparable to
$q_z$=1.6\,${\rm \AA}^{-1}$, where the quasi-Bragg surface layering peak
is expected to be observed. However, in the following paragraph
we show that as a result of resolution effects,
the limit at which the specular cusp can no
longer be distinguished from the background in practice is even lower.

The impact of the finite resolution can be easily seen by considering
a simple example of a rectangular resolution function of
infinite length along the x axis but with a very small width along the y axis.
For small values of $\alpha$ this is often an excellent approximation to
the real resolution functions \cite{Mitr01}.
With this approximation, the integration
over the $q_x$ component yields a $1/ {q_y}^{1-\eta}$ dependence for
the diffuse scattering, placing the value of
$\eta$ where the specular ridge disappears to
be $\eta$=1 instead of $\eta$=2 . For K, this $\eta$=1 limit is
reached at $q_z$=1.2\,${\rm \AA}^{-1}$.
The detector slit dimensions
which provide an optimal compromise between intensity and resolution
were discussed by one of us \cite{Pershan00} and are
$\left( width \right) \approx \left( height \right) \times \sin \alpha$, where $\alpha$ is the incident angle.

The third practical problem in extending x-ray reflectivity
measurements to large $q_z$ is the fast fall-off of
the reflected intensity.
Much more serious is the Debye-Waller-like factor that was discussed
above.
Since the surface tension $\gamma$ for K (110 mN/m at 70 $^{\circ}$\,C)
is much lower than that of a metal like Ga (770 mN/m) \cite{Iida93}, the
factor of $\exp [-\sigma^2 q_z^2]$ with $\sigma^2 \sim 1/\gamma$
is orders of magnitude smaller than that of Ga
at the same values of $q_z$. Even at high-flux synchrotron
facilities such as the Advanced Photon Source the low count rates of the reflectivity
signal observed at $q_z \approx 1.0-1.1 {\rm \AA}^{-1}$ are
among the dominant limiting factors of the $q_z$ range accessible
in this experiment.

\subsection*{Measurements}

To partially compensate for the low intensity the off-specular
diffuse scattering measurements were carried out using a linear
position sensitive detector (PSD), rather than a point detector.
In Figure 3 we show a diffuse scattering profile as a function of
$q_{xy}$.
\begin{figure}[!]
\label{fig:3}
\begin{center}
\unitlength1cm
\begin{minipage}{8.5cm}
\epsfxsize=8.5cm \epsfysize=8cm \epsfbox{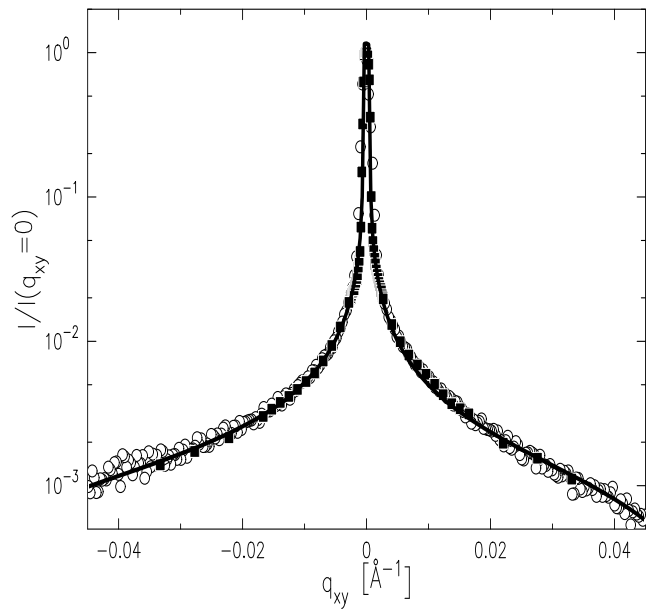}
\end{minipage}
\caption{ Comparison of an X-ray diffuse scattering scans taken
with point detector - filled squares, and position sensitive
detector (PSD) - open circles. The specular condition corresponds
to $q_z$=.3 ${\rm \AA}^{-1}$. The resolution of the bicron
detector (.22mm vertically and 2.5mm horizontally) was chosen to
match that of a PSD. Also shown is a line which represents
theoretical modeling using the capillary-wave theory for given
resolution, temperature (70$^{\circ}$\,C) and surface tension (110
mN/m). }
\end{center}
\end{figure}

The specular condition corresponds to $q_z = 0.3 {\rm \AA}^{-1}$.
The profile exhibits a sharp peak whose width is dominated by the
instrumental resolution, however the long tails are an intrinsic
property of the capillary wave surface roughness. We note that the
data shown represents the intensity difference for data obtained
at $\Delta\Theta = 0$ and {\bf $\Delta\Theta = \pm
\Delta\Theta_{offset}$} with either a position sensitive detector
(PSD) or with a point detector. An advantage of the PSD
configuration is that it permits the entire profile to be aquired
simultaneously for a wide range of the output angles $\beta$ for a
fixed $\alpha$. We carefully checked for possible instrumental
errors associated with the PSD (saturation or non-linear effects)
by comparing data obtained with both the PSD and a point detector.
As shown, in Fig 3, data obtained with both configurations are
indistinguishable. Further, this comparison allowed an accurate
determination of the PSD spatial resolution, about 0.22 mm
vertically.  To provide for a reasonable comparison with the point
detector, the vertical slit was set to match the resolution of the
PSD. The solid line in Figure 3 represents the theoretical
prediction for the different intensity, obtained by numerically
integrating $d \sigma / d \Omega$ in Eq. (1) over the known
resolution. The known values of the incident angle $\alpha$,
temperature T, surface tension $\gamma$, x-ray energy and detector
resolution, were used without any adjustable parameters. The
essentially perfect agreement between both experimental data sets
and the theoretical simulations strongly supports the theoretical
model for the effect of the thermally induced surface excitation
on the diffuse x-ray scattering from the surface.
\begin{figure}[!]
\label{fig:4}
\begin{center}
\unitlength1cm
\begin{minipage}{8.5cm}
\epsfxsize=8.5cm \epsfysize=8cm \epsfbox{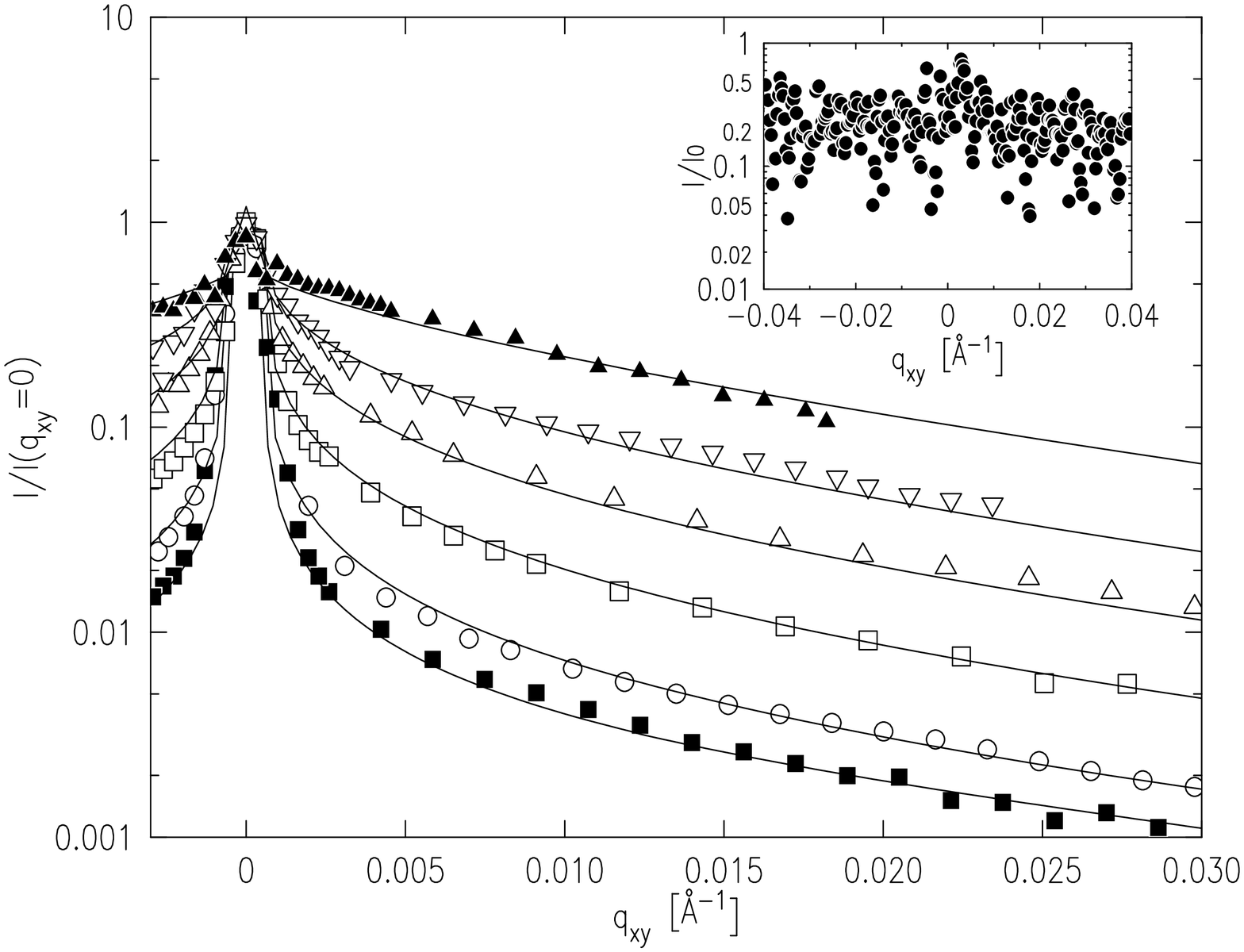}
\end{minipage}
\caption{ X-Ray diffuse scattering scans taken with PSD, as a
function of wavevector transfer component $q_{xy}$. Specular
condition corresponds to values of $q_z$ (bottom to top): 0.3,
0.4, 0.6, 0.8, 1.0 and 1.1 ${\rm \AA}^{-1}$. For comparison, the
scans are normalized to unity at $q_{xy}$=0. Lines represent
theoretically simulated scans for given values of $q_z$. The
insset shows a similar scan taken at 1.2  ${\rm \AA}^{-1}$, for
which the specular peak is marginally observable.}
\end{center}
\end{figure}

Further diffuse scattering data was collected with the PSD, which
proved to be especially useful at higher values of $q_z$, where
the weaker scattering makes it increasingly more difficult to
separate the specular signal from the power-law wings as $\eta$
approached 1. Figure 4 shows a set of diffuse scattering scans
measured with a PSD and the corresponding calculated theoretical
predictions, for several $q_z$ values (0.3, 0.4, 0.6, 0.8, 1.0 and
1.1\,${\rm \AA}^{-1}$). Each curve represents a difference between
a diffuse measurement in the plane of incidence ($\Delta\Theta =
0$) and an average of two measurements taken by displacing the
detector out of the plane of incidence by an angle of $\Delta
\Theta_{offset} = \pm$ 0.2 degrees. Theoretical simulations
include the same subtraction operation as well. All data sets are
normalized to unity at $q_{xy}$=0 for easier comparison. As $q_z$
gets larger, the ratio of specular signal to the diffuse wings
decreases, and the specular peak eventually disappears under the
off-specular diffuse wings. In the curve measured at $q_z = 1.2
{\rm \AA}^-1)$, shown in the inset to Figure 4, the specular peak
is all but disappeared. The solid lines that represent numerical
integration of the capillary-wave theory (Eq. 1) show a remarkably
good agreement with experimental data. The only adjustable
quantity in these theoretical curves is the form of $\Phi(q_z)$.
As discussed below, one common form was used for all data sets.

This agreement provides additional support for the absence of
surface contamination, since the presence of any kind of
inhomogeneous film at the surface would manifest itself in extra
x-ray diffuse signal which could not be explained by capillary
wave theory alone. Figure 5 shows the effect of varying the shape
of the detector resolution on the diffuse scattering data.
Changing the horizontal size of the slit in front of the PSD has
little effect on either the measured intensity of the specular
peak, because of the intrinsic property of the singularity, or on
the diffuse scattering at small $q_{xy}$, because it is sharply
peaked around $q_{xy} = 0$. However, far away from the specular
condition, the signal is directly proportional to the area of the
resolution function, and scales with the size of the slits.

\begin{figure}[!]
\label{fig:5}
\begin{center}
\unitlength1cm
\begin{minipage}{8.5cm}
\epsfxsize=8.5cm \epsfysize=8cm \epsfbox{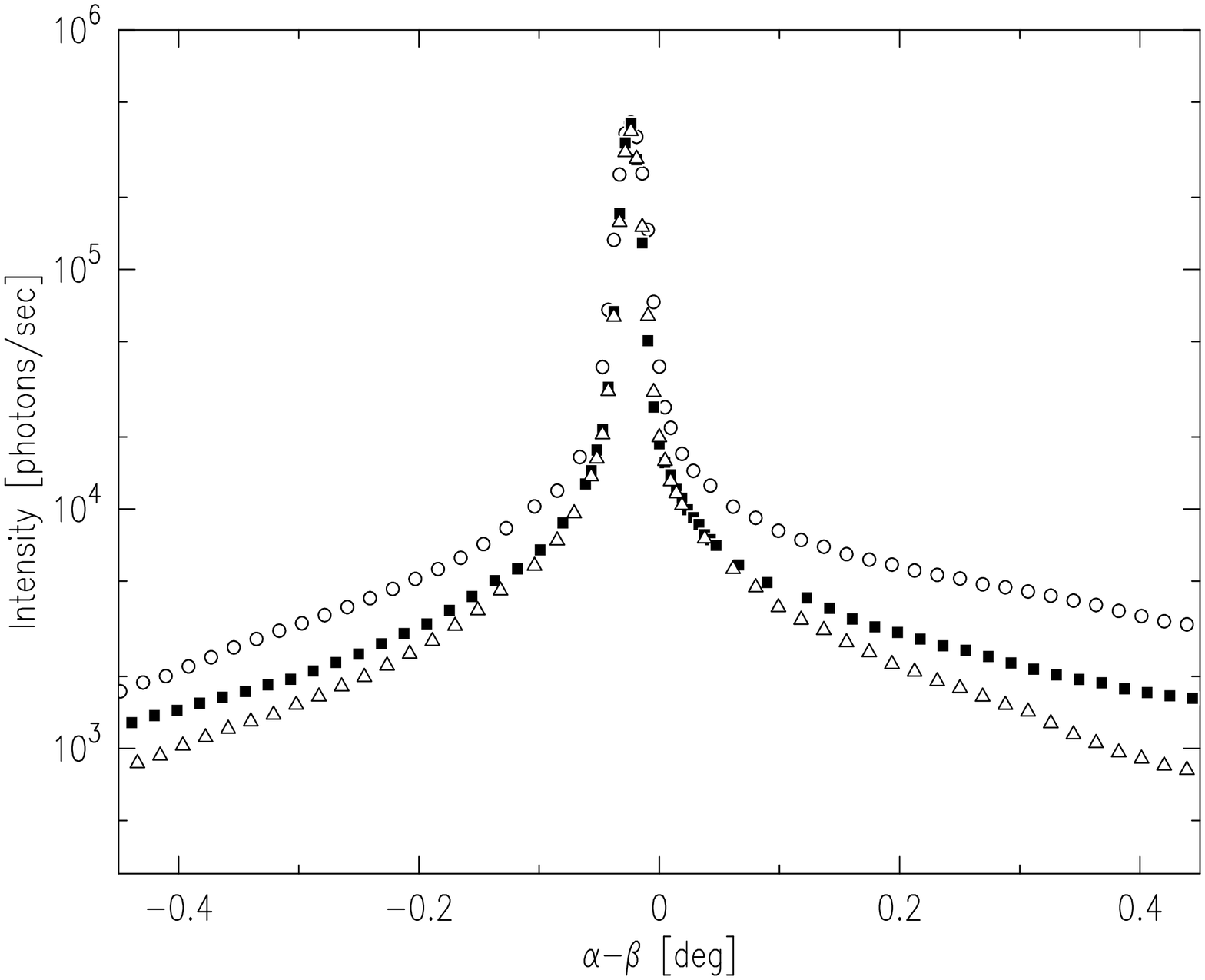}
\end{minipage}
\caption{ Effects of varying the horizontal resolution function on
the diffuse scattering data. Diffuse scans taken with PSD, peak
position corresponding to $q_z$=.4 ${\rm \AA}^{-1}$, as a function
of angular deviation from specular condition ($\alpha-\beta$) with
fixed vertical resolution (.22mm), while horizontal resolution was
changed (bottom to top: 2mm, 4mm, 8mm). }

\end{center}
\end{figure}

The excellent agreement between the experimental measurements and the
theoretical simulations strongly support our conclusion that by the
methods above one can account properly and accurately for both the
capillary-wave-induced diffuse scattering in, and the resolution effects on,
the measured data. We have, therefore, employed these results to analyze the
specular reflectivity data, to find out whether surface-induced layering
exists at the surface of K.
\begin{figure}[htb]
\label{fig:6}
\begin{center}
\unitlength1cm
\begin{minipage}{8.5cm}
\epsfxsize=8.5cm \epsfysize=8cm \epsfbox{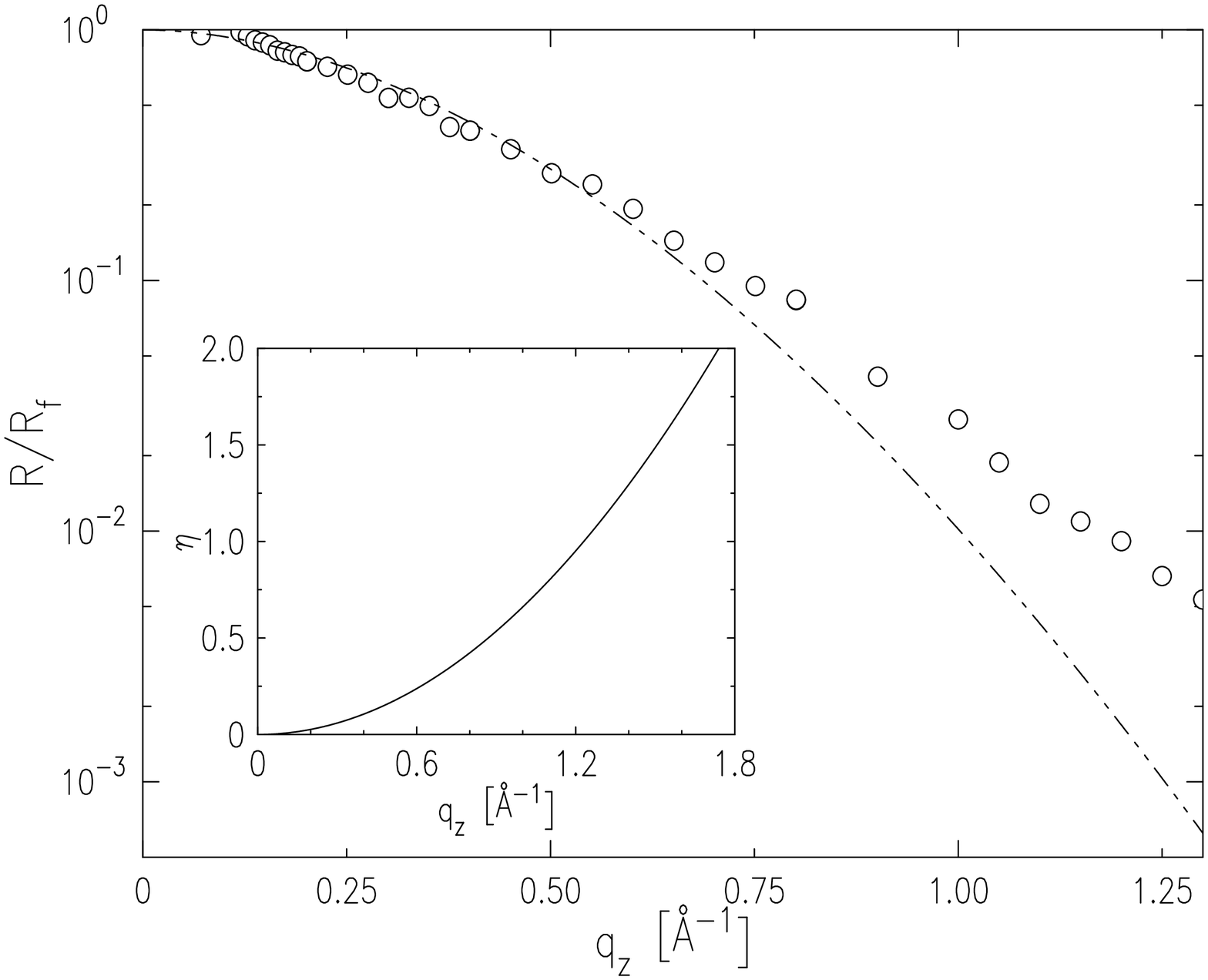}
\end{minipage}
\caption{ Fresnel-normalized reflectivity signal $R/R_f(q_z)$ for
liquid K (circles) compared to capillary wave predictions (dashed
line). The inset shows the capillary wave factor $\eta$ for liquid
K as a function of $q_z$. }
\end{center}
\end{figure}

Specular reflectivity measurements were made by scanning $\alpha=\beta$
at both $\Delta\Theta=0$ and $\Delta\Theta = \pm \Delta\Theta_{offset}$. The $\beta$-resolution,
$\Delta \beta =$ 0.2 degrees,
was set by fixing the height of the detector slit at 2.5 mm,
located 714 mm away from the center of the sample.
Likewise, the horizontal resolution was defined by a 2.5 mm
wide slit located at the same position.
The ratio of the difference between the specular signal
measured at $\Delta\Theta=0$ and at $\Delta\Theta = \pm \Delta\Theta_{offset}$
 to the theoretical Fresnel reflectivity from the K liquid-vapor
interface is shown in Fig 6. As expected,
there is a Debye-Waller type effect that causes the data to fall below unity
with increasing $q_z$. However, the measured curve (points)
decreases slower than the theoretical prediction (dashed line)
for the Debye-Waller-like factor in Eq. 1 convolved with
the resolution function.

\begin{figure}[htb]
\label{fig:7}
\begin{center}
\unitlength1cm
\begin{minipage}{8.5cm}
\epsfxsize=8.5cm \epsfysize=8.5cm \epsfbox{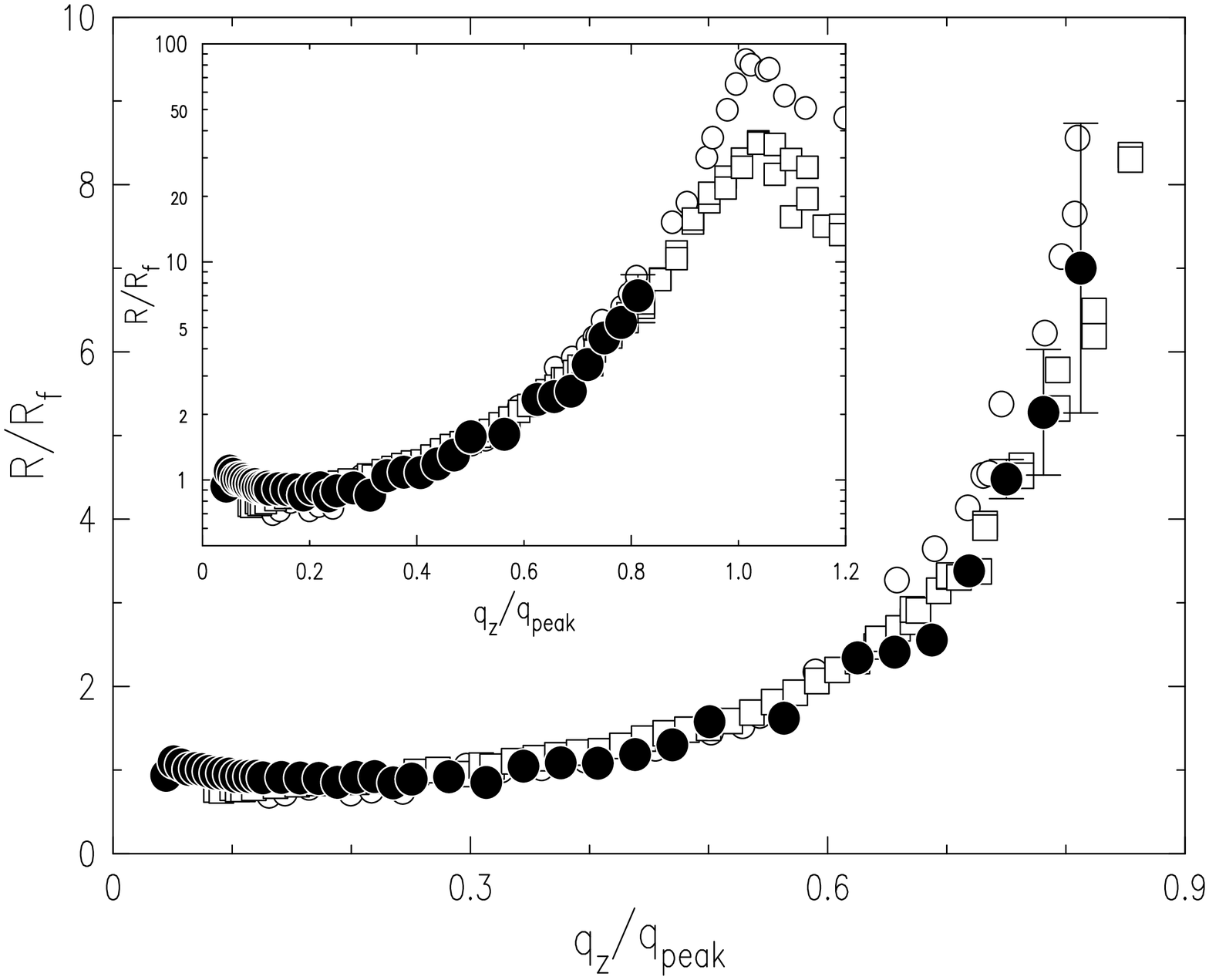}
\end{minipage}
\caption{ Surface structure factor $\Phi(q_z)$ for liquid Ga (open
circles), liquid In (open squares) and liquid K (filled circles)
obtained from X-ray reflectivity data by deconvolving resolution,
Fresnel reflectivity and capillary wave contributions, shown here
as a function of $q_z/q_{peak}$, here $q_{peak}$ is the value of
$q_z$ at which the layering peak is observed (2.4${\rm \AA}^{-1}$
for Ga and 2.2${\rm \AA}^{-1}$ for In) or is expected to be
observed (1.6${\rm \AA}^{-1}$ for K). The inset shows the same
three sets of data extended to a greater range. }
\end{center}
\end{figure}

Figure 7 shows the form factor $\Phi(q_z)$, obtained by dividing
the data in Figure 6 by the integral of the Debye-Waller factor
(see Eq. 1) over the resolution function. In order to compare it
with the surface structure factors measured previously for other
liquid metals, we have normalized $q_z$ by the value of $q_{peak}$
- the value of $q_z$ at which the layering peak is observed or is
expected to be observed. In fact we are comparing the electron
density structure factors of the different metals. Alternatively
we might have compared  the atomic densities; however, none of the
atomic form factors vary by more than a few percent over the
measured range of $q_z$. The surface structure factor of the pure
K is found here to rise clearly above unity and shows a similar
functional behavior to that of the pure Ga and In, metals for
which the surface-induced layering has been well-documented
\cite{Regan95, Tost99}. The inset shows the same plot comparing
the surface structure factors extracted from the specular
reflectivity measurements of pure Ga, pure In and pure K over an
extended $q_{z}/q_{peak}$ range. Even though experimental and
intrinsic limitations (such as the "$\eta$ effect" discussed
above) prevents us from resolving the layering peak in its'
entirety, it seems clear that liquid K exhibits surface layering
similar to that observed in other metals. These experimental
results are supported by computer simulations predicting a
surface-induced layering in pure potassium \cite{Chekmarev98}.

%\newpage
\subsection*{Discussion}

In this paper we have expressed the x-ray scattering from the
liquid surface, both specular and diffuse, in terms
of the product of a thermal fluctuation term and a surface
structure term (Eq.  3). This is analogous to the usual treatment
of Bragg scattering from three-dimensional crystals for which
the intensity  is expressed as a product of a structure factor
and a phonon-induced Debye-Waller factor. For the liquid surface
we have assumed that the thermal fluctuations could be described
by capillary waves over all length scales extending from a distance
$\xi \approx \pi/q_{max}$ that
is of the order of the  atomic or molecular size, to
a macroscopic distance that is determined by the reciprocal
of the reflectometer resolution. The gravitationally determined
capillary length  is ignored since it is many orders of magnitude
longer than the resolution-determined length.  In fact the capillary
wave model is strictly true only for long wavelengths and the
assumption that it can be invoked for lengths as small as $\xi$ is
only justified empirically  \cite{Mag95, Regan95, Tost99, Tost98, Dimasi00}. However,
the good agreement between the measured diffuse scattering
and that predicted by Eq. 1, found for a number of liquids
including the present one provides a rather general justification
for adopting this limit.

In the present study we generalized the methods that were
previously used for liquid surfaces by simultaneously carrying
out detailed measurements along and transverse to the specular
axis. As in these earlier studies we have been able to verify
the $1 / q_{xy}^{2-\eta}$ behaviour of the capillary wave form of the surface fluctuations
to values of $\eta = 0.82$ and to obtain a measure of the surface structure
factor. Aside from recent work by Mitrinovic, Williams and
Schlossman \cite{Mitr01} which were carried out to $\eta = 0.62$, the largest
values of $\eta$ attained in any of the  previous studies on liquids
were less than half of our value. In view of
the fact that the measured value of  $\Phi(q_z)$ depends on proper
isolation of the capillary diffuse scattering our confirmation
of the capillary model for diffuse scattering from  K is  essential.

The most significant limitation to the present measurement came
about as a result of the fact that due to the thermally
excited capillary wave
Debye-Waller-like effect, the  largest value of $q_z$ at
which the specular peak can be observed is smaller
than $1.6\,{\rm \AA}^{-1}$ at which  $\Phi(q_z)$ is expected
to have a peak.
This limitation can be expressed in terms of the capillary model of Eq. 1 since as
$\eta \rightarrow 2$ the specular peak no longer exists.
Another way to see the same thing is
that the scattering at the peak of  $\Phi(q_z)$ will be difficult to
observe when the product $\sigma^2(q_{peak})q_{peak}^2 \gg 1$.
In essence, this condition occurs when the effective surface roughness
approaches that of the surface layer spacing.
Although the precise value of
$\sigma(q_{peak})$ depends on the resolution, for the resolutions used
to study K, $\sigma^2(q_{peak})q_{peak}^2 = 13.8$.
Nevertheless, in this paper we demonstrated  by careful measurement
of the diffuse scattering with a number of different resolution
configurations that it was possible to quantitatively extract
values of  $\Phi(q_z)$ for K over a significant range of $q_z$. These
measurements establish that over the measured range the $q_z$-normalized
surface structure factor for pure liquid K is nearly identical
to that found in Ga and In, with a well defined rise above unity
as the peak in $q_z$ is approached from below. This strongly suggests
that the layering in K is identical to that of other liquid metals,
as implied by Chacon et al. \cite{Velasco02}.
Finally, we note that the present method could also
be applied to non-metallic liquids having similar
surface tensions \cite{Chacon01, Soler01, Chapela77}. The unresolved and intriguing question
of whether the local surface layering in metallic systems is different
from that of non-metallic liquids when the surface tensions
involved are similar will have to await future studies.

\subsection*{Acknowledgements}
This work has been supported by the U.S. Department of Energy
Grant No. DE-FG02-88-ER45379, the National
Science Foundation Grant No. DMR-01-12494 and the U.S.--Israel
Binational Science Foundation, Jerusalem. Brookhaven National
Laboratory is supported by U.S. DOE Contract No. DE-AC02-98CH10886.
PH acknowledges support from the Deutsche Forschungsgemeinschaft.

The help of Arthur Kahaian and Christopher Johnson (CMT Division,
Argonne National Laboratory) in preparing the liquid alkali metal
sample, as well as beamline assistance from Thomas Gog and
Chitra Venkataraman (CMC-CAT sector, Argonne National Laboratory)
are greatly appreciated.

\newpage
\subsection*{References}
\setlength{\parindent}{0cm}
\renewcommand{\baselinestretch}{0.8}
%\begin{small}

\end{document}